\documentclass[10pt,conference]{IEEEtran}
\usepackage{amsmath,amssymb,amsbsy,amsthm,fullpage}

\let\epsilon\varepsilon
\let\phi\varphi

\let\epsilon\varepsilon

\newtheorem*{lemma*}{Lemma}
\newtheorem{theorem}{Theorem}

\usepackage{cite}
\usepackage[pdftex]{graphicx}
\usepackage{amsmath}
\usepackage{algorithmic}
\usepackage{array}
\usepackage{caption} 
\begin{document}
\IEEEoverridecommandlockouts

\title{Using data-compressors for statistical analysis of problems on 
homogeneity testing and classification }

\author{\IEEEauthorblockN{Boris Ryabko}
\IEEEauthorblockA{Institute of Computational Technologies of SB RAS\\ Novosibirsk, Russian Federation \\}
\and
\IEEEauthorblockN{Andrey Guskov}
\IEEEauthorblockA{The State Public Scientific Technological\\ Library of SB RAS, Novosibirsk, Russian Federation \\}
\and
\IEEEauthorblockN{Irina Selivanova}
\IEEEauthorblockA{The State Public Scientific Technological \\ Library of SB RAS, Novosibirsk, Russian Federation \\}
}

\maketitle

\begin{abstract}
Nowadays  data compressors are applied to many problems of text analysis, but many such applications are 
developed outside of the framework of mathematical statistics. 
In this paper we  overcome this obstacle and show
how   several   methods of classical mathematical statistics can be  developed based  on applications of the data compressors.
 \end{abstract}


\textbf{keywords:} 
 data compression, hypothesis testing, homogeneity test, classification, universal code.






\section{Introduction}
\label{}

Data compression methods (or universal codes)
 were discovered in the 1960's  and nowadays they are widely used to compress  texts  for their storage or transmission. 
 In the last thirty  years, it was recognized that data compressors can be used for many purposes which are far from file compaction. In particular,
it was shown that methods of data compression can be used  for prediction  and hypothesis testing for time series  
in the framework of classical mathematical statistics, see  \cite{ryabko2016compression} and  a review there.
Later, several authors  applied data compressors to problems which are close, in spirit, to homogeneity testing, estimation of correlation and covariance,
classifications, clustering and some others; see \cite{kukushkina2001using,Cilibrasi:05,cilibrasi2004algorithmic,
li2004similarity,teahan2003using}. The main idea of 
their approach can be understood from the following example. Suppose that there are  three sequences of letters $x_1 x_2 ... x_n$,  $y_1$ $ y_2$ $ ...$ $ y_k$,
 $z_1 z_2 ... z_m$  and a certain data-compressor $\varphi$. 
 The sequences $x_1 x_2 ... x_n$,  $y_1$ $ y_2$ $ ...$ $ y_k$ obey  different probability distributions, whereas  $z_1 z_2 ... z_m$  obeys one of them.
 The goal is to determine this distribution. (It is the well-known ``three samples problem''.) 
 If  $x_1 x_2 ... x_n$ 
and $z_1 z_2 ... z_m$ obey the same probability distribution, then, 
the sequence $z_1 z_2 ... z_m$ will be compressed better after $x_1 x_2 ... x_n$  than after $ y_1$ $ ...$ $ y_k$. 
 More precisely, if one 
compresses sequences
$x_1 x_2 ... x_n$, $y_1$ $ y_2$ $ ...$ $ y_k$ and combined ones  $x_1 x_2 ... x_n  z_1 z_2 ... z_m$ and $y_1$ $ y_2$ $ ...$ $ y_k z_1 z_2 ... z_m$,
the difference 
$ |\varphi(x_1 x_2 ... x_n  z_1 z_2 ... z_m)| - |\varphi(x_1 x_2 ... x_n)| $
will be less than 
 $ |\varphi(y_1 y_2 ... y_k  z_1 z_2 ... z_m)| - |\varphi(y_1 y_2 ... y_k)| $
(Here $|U|$ is the length $U$.) For instance, let $x_1 x_2 ... x_n$, $z_1 z_2 ... z_m$ be texts in English, whereas  $y_1$ $ y_2$ $ ...$ $ y_k$ 
is in German. Then  the English text 
$z_1 z_2 ... z_m$ will be compressed better after the text in the same language ($x_1 x_2 ... x_n $) than after the text in German
$(y_1 y_2 ... y_k)$, i.e. the first difference will be less than the second one.

This natural approach was used for  diagnostic of the
authorship of literary and musical texts, for estimation of closeness of DNA
sequences, construction of phylogenetic trees and many other problems (\cite{kukushkina2001using,Cilibrasi:05,cilibrasi2004algorithmic,
li2004similarity,teahan2003using,vitanyi2011information,ferragina2007compression}).  
Many papers (see     
\cite{li2004similarity} and review there) were devoted to the measurement  of the interdependence between sequences
(or the  association, similarity, closeness, etc.), because such measures plays an important rule in clustering, classification and some other methods 
of text analysis. 
It is important to note, that their approaches are outside of the framework of mathematical statistics and, in particular, do  not 
give a possibility to reason about 
consistency of  estimates, tests, classifiers, clustering, etc.

It is important that the modern data compressors are based on 
so-called universal codes and the main properties of universal codes are valid for them (as far as asymptotic properties can be 
valid for a real computer program). A formal definition of universal codes is given in Appendix~1, 
but here we informally note that universal codes can compress sequences generated by a source with 
unknown statistics till its  Shannon entropy, which, in turn, is a lower limit on lossless compression. 
Note that nowadays there are many classes of universal codes which are based on different ideas and showed their practical efficiency.
Out of the most popular we mention the PPM algorithm, which is used along with the arithmetic code (\cite{bell1989modeling,rissanen1979arithmetic}), 
Lempel-Ziv codes (\cite{ziv1977universal,Ziv:78}), 
Borrows-Willer transformation (\cite{burrows1994block,manzini2001analysis}) which is used along with the ``book stack'' 
(or MTF) code (\cite{ryabko1980data,ryabko1987technical}) and some others (see for review (\cite{Rissanen:84,ryabko2016compression}).  

In this paper we show how the  idea of  compression of combined  texts described above
can be used for solving problems of homogeneity testing, classification, 
and  estimation of a measure of   interdependence, or the  association.
A distinction of the suggested method from other approaches is that it belongs to the framework of mathematical 
statistics. 
\section{Definitions and problem formulations} 

First we briefly consider the main properties of so-called universal codes, 
paying the main attention to their meaning 
whereas formal definitions of codes, stationary ergodic sources and Shannon entropy are given in Appendix 1 and
can be found in \cite{Cover:06,ryabko2016compression}. 

Here we only note that we  consider so-called lossless codes $\varphi$  which  encode words over alphabet $A$ 
using words from the binary alphabet $\{0,1\}$ in such a way that any word can be decoded without mistakes, i.e. 
there exists a map $\varphi^{-1}$ such that for any
word $w$ over $A \, \,  $ $ \, \varphi^{-1}(\varphi(w)) = w$. 
Let a code $\varphi$ be applied to encode sequences generated by a stationary ergodic source
$\mu$. The value $
 R_t(\varphi, \mu) =  \frac{1}{t} \, |\varphi (x_1 ... x_t) | \, - \, h_\infty(\mu) $
is called the redundancy, where $h_\infty(\mu)$ is the limit entropy, see Appendix 1. (It is known that $h_\infty(\mu)$ is an
attainable lower limit for lossless codes, that is why  the difference is called the redundancy.)
By definition, the code $\varphi$ is called universal if the redundancy goes to 0 as $t$ grows, i.e.
$ \lim_{t \rightarrow \infty}  R_t(\varphi, \mu) = 0 \, .$
(In other words, a universal code  compresses sequences generated by any stationary ergodic  source till the limit value.)

The main goal of the paper is to give a compression-based solution for the following problems: 

i) Homogeneity test, where there are several sequences $x^1_1x^1_2 ... x^1_{n_1}$, 
 $...$, $x^k_1x^k_2 ... x^k_{n_k}$,  $y^1_1 y^1_2 ... y^1_{m_1}$,   $...$
 $y^s_1 y^s_2 ... y^s_{m_s}$, generated either by  a single  source or by two different ones, and two corresponding  hypotheses. 
 We also consider the more general case  where there are more than two different sets of sequences.
 
 ii) Classification problems, where there are   samples  $x^1_1x^1_2 ... x^1_{n_1}$, 
 $...$, $x^k_1x^k_2 ... x^k_{n_k}$,  $y^1_1 y^1_2 ... y^1_{m_1}$,   $...$
 $y^s_1 y^s_2 ... y^s_{m_s}$ generated two different (but unknown) sources and $z_1 ... z_l$ is generated by one of the two. The goal is to 
 determine which of them generated  $z_1 ... z_l$.
 
 iii)  Estimation of a so-called 
 measurement  of  interdependence, or the  association. 
  \section{The main theorems}
 First we present a theorem which can be considered as a theoretical basis for application of data compressors
for solving the problems described above. 
First, for two words $v_1 .. v_l$ and $u_1 .. u_s$ we define 
the following value:
\begin{equation}\label{dif} 
|\varphi(v_1 .. v_l/u_1 .. u_s)| = |\varphi(u_1 .. u_s v_1 .. v_l)|  - |\varphi(u_1 .. u_s)|, 
\end{equation}
where  $s,l$ are integers.
Informally, it is the length of the codeword for $v_1 .. v_l$ if it is encoded with the word $u_1 .. u_s$.
 \begin{theorem}\label{mai}
 Let $\mu_x$ and $\mu_y$ be stationary ergodic sources generating letters from a finite alphabet $A$ and let
 their  memory be upper-bounded by a certain constant $M$. Suppose that   $ ... y_{-t} $ $y_{-t+1}$ $ ...$ $ y_{-1}y_0$ is generated by $\mu_y$,
 whereas 
  $... x_{-t} x_{-t+1} ... x_{-1} x_0$ and $  w_1 ... w_m, m \ge 1,  $ are
 generated by $\mu_x$ in such a way that all three sequences are independent,  and let $\varphi$ be a universal code. Define 
   \begin{equation}\label{del}    
  \Delta_{t,k,m}  = 
  |\varphi(w_1 ... w_m/ y_{-k} y_{-k+1} ... y_{-1}y_0)| - $$ $$ |\varphi(w_1 ... w_m/  x_{-t} x_{-t+1} ... x_{-1} x_0) |
   \end{equation}
  where $ |\varphi(v_1 .. v_l/u_1 .. u_s)|  $ is defined in (\ref{dif}).
  Then
  there exists such a constant $\lambda $ and an integer $m_0$  that for any  $m > m_0$ there is 
   such an integer $L$ that, for any $t > m + L$, $k > m + L$,
  \begin{equation}\label{main}
E_{\mu_x} E_{\mu_y} (\Delta_{t,k,m} ) \, \ge \, \lambda \, m +\circ(m) \,  ,
  \end{equation} and, if $\mu_x = \mu_y$ then 
  $\lambda = 0$, otherwise $\lambda > 0$. Here      
  $E_\nu(  )$ is the  expectation with respect to the measure $\nu$. 
  \end{theorem}
    The proof is given in Appendix 2, whereas here we give some explanations of the  theorem. 
    The theorem compares two cases:   
    the word $w_1 ... w_m$ is compressed either together with  $x_{-t} x_{-t+1} ... x_{-1} x_0$ or with $y_{-k} y_{-k+1} ... y_{-1}y_0$.
  The word $w_1 ... w_m$  is generated by $\mu_x$, hence, it should be compressed better with  $x_{-t} x_{-t+1} ... x_{-1} x_0$ and the value
  $|\varphi(w_1 ... w_t/  x_{-t} x_{-t+1} ... x_{-1} x_0) | ( = $ $|\varphi( x_{-t} x_{-t+1} $ $... $ $ x_{-1} x_0w_1 ... w_t)| -$ $ 
  |\varphi( x_{-t} x_{-t+1} ... x_{-1} x_0) | ) $ should be less than  $|\varphi(w_1 ... w_t/ y_{-k}$ $ y_{-k+1} ... y_{-1}y_0)| $.
  The theorem shows, that, indeed, the word $w_1 ... w_m$ is compressed  better together with  $x_{-t} x_{-t+1} ... x_{-1} x_0$
  if the lengths of sequences $w_1w_2 ... w_m$, $x_{-t}$ $  x_{-t+1}$ $  ... x_{-1} x_0$ and $y_{-k} y_{-k+1} ... y_{-1}y_0$ are sufficiently large.

   
  
Let there be  two  sets of sequences: $X= \{$ $x^1_1x^1_2 ... x^1_{n_1}$, 
$x^2_1x^2_2 ... x^2_{n_2}$, $...$, $x^k_1x^k_2 ... x^k_{n_k} \}$, and $Y = \{$  $y^1_1 y^1_2 ..., y^1_{m_1}$,  $y^2_1 y^2_2 ... y^2_{m_2}$, $...$
 $y^s_1 y^s_2 ... y^s_{m_s} \}$, generated 
 by possibly different measures $\mu_x$ and $\mu_y$. We consider two hypotheses $H_0 = \{ \mu_x = \mu_y \}$ and $H_1 = \{ \mu_x \neq \mu_y \}$
 and our goal is to develop a statistical test for them using the sets $X$ and $Y$. 
 First we give an informal description   of the suggested test, which will be based on data compression.
 Combine 
  a half of the  sequences from the set  $X$ into  $X^*$ (say,  $x^1_1x^1_2 ... x^1_{n_1}$, 
  $x^2_1x^2_2 ... x^2_{n_1}$, ... , $x^{\lfloor k/2 \rfloor}_1x^{\lfloor k/2 \rfloor}_2 ... x^{\lfloor k/2 \rfloor}_{n_{\lfloor k/2 \rfloor}}$ ) 
  and  half of  $Y$ (say  
  and $y^1_1 y^1_2 ... y^1_{m_1}$, $y^1_1 y^2_2 ... y^2_{m_2}$, $... ,$  $y^{\lfloor s/2 \rfloor}_1 y^{\lfloor s/2 \rfloor}_2 ... 
  y^{\lfloor s/2 \rfloor}_{m_{\lfloor s/2 \rfloor}}$)) into $Y^*$. Then
   compress  all other sequences using  a universal code $\varphi$ along with $X^*$ and $Y^*$. If $H_0$ is true, the values 
  $ |\varphi (x^i_1x^i_2 ... x^i_{n_i} /$ $ X^*)| \, ,$
  $i= \lfloor k/2 \rfloor +1, ..., k$ and 
 $ |\varphi (y^j_1 y^j_2 ... y^j_{m_j} / $ $ X^*)| \, ,$ 
 $j=\lfloor s/2 \rfloor + 1, ..., s$ should be  evenly mixed. Otherwise, if $H_1$ is true, then, 
 on average, the numbers from the second set should be larger than those from the first one,
 because the probability distributions of the  sequences from
 $y^i_j y^i_j ... y^i_{m_j}$ and $X^*$ are different, whereas the
 probability distributions of the  sequences 
 $x^i_1x^i_2 ... x^i_{n_i}$ and $X^*$ are the same (here and below $|U|$ is the number of elements in $U$). 
 
 {\it The test}. A more formal description of the suggested test is as follows: 
 i) Denote the set $X \setminus X^* $ by $\hat X$ and  $Y \setminus Y^* $ by $\hat Y$
 and calculate  for any $x^i_1x^i_2 ... x^i_{n_i}$ from $\hat X$ the  values 
 \begin{equation}\label{ga}
  \gamma_i =|\varphi (x^i_1x^i_2 ... x^i_{n_i} / Y^* )| \,- 
 \, |\varphi (x^i_1x^i_2 ... x^i_{n_i} / X^* )| 
 \end{equation}
 and
 for any $y^j_1 y^j_2 ... y^j_{m_j}$ from $\hat Y$
  \begin{equation}\label{de}
    \delta_j =|\varphi (y^j_1 y^j_2 ... y^j_{m_j}  /  X^*)| \,- \, 
  |\varphi ( y^j_1 y^j_2 ... y^j_{m_j} /  Y^* )| \, .
  \end{equation}
 Define 
 \begin{equation}\label{nn} 
  n_{1,1} = |\{i : \gamma_i \ge 0 \} |, \, n_{1,2} = |\{ i : \gamma_i< 0 \} |, $$ $$
  n_{2,1} = |\{i : \delta_i <0 \} |, n_{2,2} = |\{ i : \delta_i  \ge 0 \} | \, .
 \end{equation}

 ii) Apply the test   of the independence for the $2 \times 2$ table to
 
\[N_{2,2} = \left| \begin{array}{cc}
 n_{1,1}  & n_{1,2}   \\
n_{2,1}  & n_{2,2} \\
\end{array} \right|.\] 
  A detailed analysis of this problem is carried out in \cite{Kendall:61}, part 33. In particular, there is a
  description of efficient tests for  homogeneity problem for the $2 \times 2$ table (see part 33.22).
  We denote this test as $\Psi_\alpha$, where $\alpha$ is the level of significance. Note, that there 
  are some requirements for   values of $n_{1,1},  n_{1,2},  n_{2,1},  n_{2,2} $ which should be valid if $\Psi_\alpha$ is used (see \cite{Kendall:61}). 
\begin{theorem}\label{hom}
Let $\mu_x$ and $\mu_y$ be stationary ergodic measures whose memory is finite (but, possibly, unknown).
 If the above described test is applied for testing $H_0$ against $H_1$ along with the test
 $\Psi_\alpha$ and the requirements of  $\Psi_\alpha$ for   values of $n_{1,1},  n_{1,2},  n_{2,1},  n_{2,2} $ are valid,
 then for any code $\varphi$
 the Type I error is not grater than $\alpha$. 
 
 If  $\varphi$ is a universal code, $|X|$ and $|Y|$ go to infinity and the 
 lengths of all sequences from $X$ and $Y$ go to infinity in such a way that for all 
 sequences $x, x^* \in X$, $y, y^* \in Y$   the ratios $|x| / |x^*|$ and  $|y| / |y^*|$ are upper-bounded by a certain constant, then, 
 with probability 1, 
 the Type II error goes to 0. 
\end{theorem}
The proof is given in Appendix 2, but here we give some comments.
 First, note that there are many tests of homogeneity for $2 \times 2$ tables and, in principle,  any of 
 them can be used. That is why, we do not describe the test 
 $\Psi_\alpha$ on   sizes of $n_{1,1},  n_{1,2},  n_{2,1},  n_{2,2} $. 
 Second,  the described method can be  easily  extended from the two-sample problem  to the $s$-sample problem, $s >2$.
Namely, let there be $s, s>2,$
sets  $V_1$, $V_2, ...,$ $V_s$ of sequences generated by stationary ergodic sources
$\mu_1, \mu_2$ $  ...,  \mu_s$. Let there be
the hypotheses $H_0 =$ $\{\mu_1 = \mu_2$ $ = ... = \mu_s \} $ and $H_1 = \bar H_0$.
In this case we carry out calculations based on the scheme  described above in order to obtain a
so-called $s \times s$ table and then apply a test for homogeneity, see    (\cite{Kendall:61}).  
 
{\it Measurement  of the interdependence and association.} 
If the  hypothesis of homogeneity is rejected, it is natural to measure interdependence.
We suggest to measure interdependence between two sets of sequences $X$ and $Y$ (and the corresponding sources) based on 
the above described  $2 \times 2$ tables. (In the case of the $s$-sample problem, $s >2$, the measures will be based on the $s \times s$ table.) 
This problem is well-investigated in the mathematical statistics, see, for example,   (\cite{Kendall:61}, part 33). 
That is why, we mention such measures only briefly. 
For $2 \times 2$ tables we mention the coefficient of association, $ Q$, defined by
the equation
$
Q = (n_{1,1} n_{2,2} - n_{1,2} n_{2,1})/n_{1,1} n_{2,2} + n_{1,2} n_{2,1})
$
and the coefficient
$
V =   (n_{1,1} n_{2,2} - n_{1,2} n_{2,1})/ $ $ \sqrt{ (n_{1,1}+ n_{2,2}) ( n_{1,1}+ n_{2,1}) ( n_{1,2}+ n_{2,2}) ( n_{2,1}+ n_{2,2})   },
$
see,   (\cite{Kendall:61}, part 33). It is important to note that there are well-known methods of building  standard errors and confidence interval 
 for $Q$ and $V$ (\cite{Kendall:61}, part 33).

  Let there be sequences $\hat w^1 = w^1_1 w^1_2 ... w^1_{m_1}$, $\hat w^2 = w^2_1 w^2_2 ... w^2_{m_2}$, 
$... ,$ $\hat w^k = w^k_1 w^k_2 ... w^k_{m_k}$, generated by stationary ergodic sources $\nu_1$, $ ..., $ $ \nu_k$, correspondingly, 
where $k \ge 2$. There is a new sequence $\hat u = u_1u_2 ... u_n$, $n \ge 2$, and it is known beforehand that it is generated by one of 
the sources from $\{ \nu_1$, $ ..., $ $ \nu_k \}$. The problem of classification is to determine which source generated the sequence  
$ \hat u = u_1u_2 ... u_n$.
By definition, a method of classification  is called asymptotically  consistent
if,
with probability 1, the method finds $\nu$ which generated the sequence $u_1u_2 ... u_n$ when $\min(n, m_1, m_2, ..., m_{k})$ goes to infinity.

{\it Method of classification.} We suggest the following method of classification: decide that the sequence 
$u_1u_2 ... u_n$ was generated the source $\nu_j$ for which
\begin{equation}\label{cl}
 j =   \arg \min_{i=1, ..., k} |\varphi( u_1u_2 ... u_n/ w^i_1 w^i_2 ... w^i_{m_k})| \, .
\end{equation}
\begin{theorem}\label{tclas}
Let $\nu_1$, $ ..., $ $ \nu_k$ be stationary ergodic measures whose memories are finite (but, possibly, unknown). If 
$\varphi$ is a universal code and the lengths of all sequences $\hat u, $ $ \hat w_1,$ $ \hat w_2,$ $  ..., $ $\hat w_k $ go to infinity in such a way 
 that   
 \begin{equation}\label{cC}
 \lim_{ |\hat u|  \rightarrow \infty}  |\hat u|/ \min_{j} |\hat w^j| \, = \, 0 \, ,
 \end{equation}
 then
    the described method is asymptotically  consistent.
\end{theorem}

\section{Appendix 1}
Let $\tau$ be a stationary  ergodic source generating letters
from a finite alphabet $A$. (Definitions can be found in \cite{Cover:06}.) The $m-$ order (conditional) Shannon
entropy and the limit Shannon entropy are defined as follows:
\begin{equation}\label{moe} h_m(\tau) = \, - \,
\sum_{v \in A^m} \tau(v) \sum_{a \in A} \tau(a|\,v) \log
\tau(a|\,v),  $$ $$
 \: h_\infty(\tau) = \lim_{m \rightarrow \infty} h_m(\tau). \end{equation}
 It is known that for any integer $m$
 \begin{equation}\label{ent-ineq} h_m(\tau) \ge  h_{m+1}(\tau)
 \, \ge h_\infty(\tau),  \end{equation} see 
\cite{Cover:06}.  
Now we define codes. Let $A^\infty$ be
the set of all infinite words $x_1x_2 \ldots $ over the alphabet
$A$. A data compression method (or code) $\varphi$ is defined as a
set of mappings $\varphi_n $ such that $\varphi_n : A^n
\rightarrow \{ 0,1 \}^*,\, n= 1,2, \ldots\, $ and for each pair of
different words $x,y \in A^n \:$ $\varphi_n(x) \neq \varphi_n(y)
.$ Informally, it means that the code $\varphi$ can be applied for
compression of each message of any length $n$ over the alphabet $A$
and the message can be decoded if its code is known. It is also
required that each sequence $\varphi_n(u_1)\varphi_n(u_2)
...\varphi_n(u_r), r \geq 1,$ of encoded words from the set $A^n,
n\geq 1,$ can be uniquely decoded into $u_1u_2 ...u_r$. Such codes
are called uniquely decodable. For example, let $A=\{a,b\}$, the
code $\psi_1(a) = 0, \psi_1(b) = 00, $  obviously, is not uniquely
decodable. It is well known
 that if  a code $\varphi$ is uniquely decodable
then  the lengths of the codewords satisfy the following
inequality (Kraft inequality): $ \Sigma_{u \in A^n}\: 2^{-
|\varphi (u) |} \leq 1\:,$ see, for ex., \cite{Cover:06}. 
Moreover, if the sum $ \Sigma_{u \in A^n}\: 2^{-
|\varphi (u) |}$ is less than 1, there exists such a code $\varphi^*$ that 
 i) $ \Sigma_{u \in A^n}\: 2^{-
|\varphi^* (u) |} = 1\:,$ and 
$ |\varphi^* (u) | \le |\varphi (u) |$ for any word $u$, \cite{Cover:06}. (Informally, it means, that $\varphi^*$ compresses better.)
So, we can consider only codes for which 
\begin{equation}\label{kraft}
 \Sigma_{u \in A^n}\: 2^{-
|\varphi (u) |} = 1\: . 
\end{equation}
We will use a so-called  Kullback-Leibler (KL) divergence, which is
defined by
\begin{equation}\label{kl}
D(P||Q) = \sum_{b\in B} P(b) \log \frac{P(b)}{Q(b)} \: ,
\end{equation} where $P(b)$ and $Q(b)$ are probability
distributions over  alphabet $B$.
It is known that
for any distributions $P$ and $Q$ the KL divergence is nonnegative
and equals $0$ if and only if $P(a) = Q(a)$ for all $a \:$.

Let us  describe universal codes, or data compressors. (A 
detailed description can be found, for example, in \cite{ryabko2016compression}.)
First we note that (as
 is known in Information Theory) sequences 
generated by a  source $p$ can be "compressed" on average  till 
the limit Shannon entropy $h_\infty(p)$  and, on the other hand, there is no code $\psi$ for which the expected codeword
 length is less than  $h_\infty(p)$; see \cite{Cover:06}. 
  As  defined above, a code $\varphi$ is universal if it compresses till this limit.
 We will consider universal codes which can be applied to any word $x_1 ... x_s$, $n>0$, from a certain alphabet $A$ 
 and such that the following natural property is valid:  
 for any words $u$, $v, |v| >0, $ 
 \begin{equation}\label{nonze}
 | \varphi(uv)| - |\varphi(u)| >0 \, .
 \end{equation} 
For any universal code $\varphi$  we define a measure $\pi_\varphi$ as follows:
\begin{equation}\label{pi}
 \pi_\varphi (x_1x_2 ... x_t) \, = \, 2^{ - |\varphi (x_1x_2 ... x_t)|} \, ,
\end{equation}
where $x_1x_2 ... x_t$ is a word over $A$. It is  known that for any $t$ 
\begin{equation}\label{sum} 
 \sum_{u \in A^t} \pi_\varphi (u) \, = \, 1 \, ,
\end{equation}
(see the Kraft inequality and (\ref{kraft})).

Now we consider an application of universal codes (or data compressors) which is the main tool for solving the  problems mentioned  in Introduction.
From (\ref{dif}) and (\ref{pi}) we immediately obtain that  
\begin{equation}\label{difme} 
|\varphi(v_1 .. v_l/u_1 .. u_s)| = -  \log( \pi_\varphi(v_1 .. v_l|u_1 .. u_s) ) \,
\end{equation}
and the right part is defined correctly, see (\ref{nonze}).  
 The following property of universal codes will play an important role in what follows. 
  Let $\mu_z$ be a stationary ergodic source generating letters from an alphabet $A$ and  $\varphi$ be a universal code.
 For any integer $m \ge 1$ and 
$w_1 ... w_m \in A^m$ with probability 1
\begin{equation}\label{klaim}
 \lim_{T   \rightarrow \infty}  \log \frac{\pi_\varphi(w_1 w_2 ... w_m | z_{-T} ... z_0)}{\mu_z(w_1 w_2 ... w_m | z_{-T} ... z_0 )} = 0 \, .
\end{equation}
Note that this equation  shows that the code $\varphi$ estimates the (unknown) probability precisely, where $T$ grows.

\section{Appendix 2: proofs}

{\it Proof of Theorem 1.}
First we prove the following 

{\it Claim.}
\begin{equation}\label{mx-my}
 E_{\mu_x} E_{\mu_y} \sum_{w_1w_2 ... w_m}  \mu_x(w_1w_2 ... w_m) $$ $$ \log \frac{\mu_x(w_1... w_m|x_{-t} x_{-t+1} ... x_0) }
 {\mu_y(w_1... w_m|y_{-k} y_{-k+1} ... y_0 )}
 = (m-M)\lambda + O(1) \, ,
\end{equation}
where 
 $M$ is a upper bound of memories of $\mu_x$ and $\mu_y$, and $\lambda$ is such a constant that $\lambda = 0$ if $\mu_x = $ $\mu_y$ and
 $\lambda > 0$ otherwise. 
 
{\it Proof of the claim.}
The left side can be presented as follows:
\begin{equation}\label{mx-my2}
 E_{\mu_x} E_{\mu_y} \sum_{w_1w_2 ... w_m}  \mu_x(w_1w_2 ... w_m) $$ $$\log \frac{\mu_x(w_1... w_m|x_{-t} x_{-t+1} ... x_0) }
 {\mu_y(w_1... w_m|y_{-k} y_{-k+1} ... y_0 )} =
 $$
$$ E_{\mu_x} E_{\mu_y}  (
\sum_{w_1w_2 ... w_{m-1}} \mu_x(w_1w_2 ... w_{m-1}) $$ $$  \sum_{w_m}  \mu_x(w_m|w_1w_2 ... w_{m-1}) $$ $$  \log \frac{\mu_x(w_m|x_{-t} x_{-t+1} ... x_0w_1w_2 ... w_{m-1}) }
 {\mu_y(w_m|y_{-k} y_{-k+1} ... y_0w_1w_2 ... w_{m-1} )}
 $$
 $$ + \sum_{w_1w_2 ... w_{m-2}}  \mu_x(w_1w_2 ... w_{m-2}) $$ $$ \sum_{w_{m-1}}  \mu_x(w_{m-1}|w_1w_2 ... w_{m-2}) $$ $$
 \log \frac{\mu_x(w_{m-1}|x_{-t} x_{-t+1} ... x_0w_1w_2 ... w_{m-2}) }
 {\mu_y(w_{m-1}|y_{-k} y_{-k+1} ... y_0w_1w_2 ... w_{m-2} )} $$
 $$
+ ... + \sum_{w_1} \mu_x(w_1)     \log \frac{\mu_x(w_{1}|x_{-t} x_{-t+1} ... x_0) }
 {\mu_y(w_1|y_{-k} y_{-k+1} ... y_0)}  ) \, .
\end{equation}
It is supposed that the memory of $\mu_x$ and $\mu_y$ is upper-bounded by $M$. Hence, by definition, it means that
$\mu_r (v|u_i u_{i+1} ... u_j) $ $ = \mu_r (v|u_{j-M+1}  u_{j-M +2} ... u_j),$ where $j \ge i +M-1$.
From this equation and the previous one we obtain the following equation:
\begin{equation}\label{mx-my3}
 E_{\mu_x} E_{\mu_y} \sum_{w_1w_2 ... w_m}  \mu_x(w_1w_2 ... w_m) $$ $$ \log \frac{\mu_x(w_1... w_m|x_{-t} x_{-t+1} ... x_0) }
 {\mu_y(w_1... w_m|y_{-k} y_{-k+1} ... y_0 )} =
 $$
$$  
\sum_{w_1w_2 ... w_{m-1}} \mu_x(w_1w_2 ... w_{m-1})   \sum_{w_m}  \mu_x(w_m|w_1w_2 ... w_{m-1}) $$ $$
\log \frac{\mu_x(w_m|w_{m-M} ... w_{m-1}) }
 {\mu_y(w_m|w_{m-M} ... w_{m-1} )}
 $$
 $$ + \sum_{w_1 ... w_{m-2}}  \mu_x(w_1... w_{m-2}) \sum_{w_{m-1}} \mu_x(w_{m-1}|w_1 ... w_{m-2}) $$ $$
 \log \frac{\mu_x(w_{m-M-1} ... w_{m-2}) }
 {\mu_y(w_{m-M-1} ... w_{m-2}) } 
 $$
 $$ + ...+ \sum_{w_1... w_{M}}  \mu_x(w_1... w_{M}) \sum_{w_{M+1}} \mu_x(w_{M+1}|w_1 ... w_{M}) $$ $$
 \log \frac{\mu_x((w_{M+1}|w_1 ... w_{M}) }
 {\mu_y((w_{M+1}|w_1w_2 ... w_{M}) } $$
 $$ + E_{\mu_x} E_{\mu_y}  (
    \sum_{w_1w_2 ... w_{M-1}}  \mu_x(w_1w_2 ... w_{M-1}) $$ $$  \sum_{w_{M}} \mu_x(w_{M}|x_0w_1w_2 ... w_{M-1})
 $$ $$ \log \frac{\mu_x((w_{M}|x_0w_1w_2 ... w_{M-1}) }
 {\mu_y((w_{M}|y_0w_1w_2 ... w_{M-1}) } + ...$$ $$ + \sum_{w_1} \mu_x(w_1)     \log \frac{\mu_x(w_{1}|x_{-M+1} x_{-M+2} ... x_0) }
 {\mu_y(w_1|y_{-{M+1}} y_{-M+2} ... y_0)}  ) \, .
\end{equation}
For any function $\psi(u_{e+1}, ...,u_{e+b})$ and any measure $\nu(u_1u_2 ... u_eu_{e+1} ...u_{e+b})$
$$ \sum_{u_1u_2 ... u_eu_{e+1} ...u_{e+b}} \nu(u_1u_2 ... u_eu_{e+1} ...u_{e+b}) $$ $$ \psi(u_{e+1}, ...,u_{e+b}) $$ $$ = 
\sum_{  u_{e+1} ...u_{e+b}} \nu(u_{e+1} ...u_{e+b}) \psi(u_{e+1}, ...,u_{e+b} ) \, ,
$$
where $e$ and $b$ are integers. Having taken into account this equation, stationarity $\mu_x$ and $\mu_y$,
equations (\ref{mx-my}), (\ref{mx-my2}) and (\ref{mx-my3}), we obtain 
$$
 E_{\mu_x} E_{\mu_y} \sum_{w_1w_2 ... w_m}   \mu_x(w_1w_2 ... w_m) $$ $$ \log \frac{\mu_x(w_1... w_m|x_{-t} x_{-t+1} ... x_0) }
 {\mu_y(w_1... w_m|y_{-k} y_{-k+1} ... y_0 )} = 
 $$
$$
(m - M) \sum_{w_1w_2 ... w_{m-2}}  \mu_x(w_1w_2 ... w_{M}) $$ $$ \sum_{w_{M+1}} \mu_x(w_{M+1}|w_1w_2 ... w_{M})
$$ $$
 \log \frac{\mu_x((w_{M+1}|w_1w_2 ... w_{M}) }
 {\mu_y((w_{M+1}|w_1w_2 ... w_{M}) } 
 + O(1) \, . 
 $$
 From properties of K-L divergence (see  (\ref{kl}) ) we can see that $\lambda > 0$ if $\mu_x \neq \mu_y$ and, obviously, $\lambda = 0$,
 if $\mu_x = \mu_y$.  
 The claim is proven.

Let us proceed with proof of the theorem.
 Having taken into account the definitions (\ref{difme}) and (\ref{del}), we obtain that 
 \begin{equation}\label{chain} 
 E_{\mu_x} E_{\mu_y} (\Delta_{t,k,m} ) = $$ $$ \sum_{u_{-t}u_{-t+1} ... u_0 \in A^{t+1}} 
 \sum_{v_{-k}v_{-k+1} ... v_0 \in A^{k+1}} \sum_{w_1w_2 ... u_m \in A^{m}} $$
$$ \mu_y(v_{-k}v_{-k+1} ... v_0) \mu_x(u_{-t}u_{-t+1} ... u_0)  \mu_x(w_1w_2 ... u_m) $$  $$
\log \frac{\pi_\varphi ( w_1 ... w_m|u_{-t}u_{-t+1} ... u_0) }{\pi_\varphi ( w_1 ... w_m|v_{-k}v_{-k+1} ... v_0) } .
 \end{equation}
 The following equation is obvious:
 \begin{equation}\label{chain2} 
  \log \frac{\pi_\varphi ( w_1 ... w_m|u_{-t}u_{-t+1} ... u_0) }{\pi_\varphi ( w_1 ... w_m|v_{-k}v_{-k+1} ... v_0) }  = $$ $$
   \log \frac{\mu_x( w_1 ... w_m|u_{-t}u_{-t+1} ... u_0) }{\mu_y ( w_1 ... w_m|v_{-k}v_{-k+1} ... v_0) } +$$ $$
    \log \frac{\pi_\varphi ( w_1 ... w_m|u_{-t}u_{-t+1} ... u_0) }{\mu_x ( w_1 ... w_m|u_{-t}u_{-t+1} ... u_0) }  
   + $$ $$ \log \frac{\mu_y ( w_1 ... w_m|v_{-k}v_{-k+1} ... v_0) }{\pi_\varphi ( w_1 ... w_m|v_{-k}v_{-k+1} ... v_0) } .
 \end{equation}
 The first term is estimated in the claim, see (\ref{mx-my}), whereas the second and the third terms can be estimated based on (\ref{klaim}).
So, from the claim and (\ref{klaim}), we can see that,  with probability 1, 
$ E_{\mu_x} E_{\mu_y} (\Delta_{t,k,m} ) = (m-M) \lambda + O(1) \, .
$
The theorem is proven.

Proof of the Theorem~\ref{hom}.
First we consider the case where $H_0$ is true. It means that the sequences from $\hat{X}$ and $\hat{Y}$ obey the same distribution. Hence,
$\gamma_i$ (\ref{ga}) and $\delta_j$ (\ref{de})  have the same distribution, too, and the above mentioned test $\Psi_\alpha $
from \cite{Kendall:61}, part 33, can be applied.
Now  we consider the case where $H_1$ is true.
In this case the length of any sequence grows, so, the length will be grater than $m_0$ from Theorem~\ref{mai}. The number of sequences
grows to infinity and the total  length of a half of them goes to infinity in such a way that for any integer $L$ the total length will be grater
than the sum $m+L$ from  Theorem~\ref{mai}. From this theorem we can see that $n_{1,2}$ and $n_{2,1}$ goes to 0 and, hence,
the Type II error goes to 0.

Proof of the Theorem~\ref{tclas}.
Suppose, that the sequence $u_1u_2 ... u_n$ was generated by $\nu_j$. Then, we can see from Theorem~\ref{mai} that, with probability 1,
the value $|\varphi( u_1u_2 ... u_n/  w^i_1 w^i_2 ... w^i_{m_k})| $ grows as $\lambda_{i} n + \circ (n) $, $\lambda_{i} > 0$,  if $i \neq j$, 
(i.e. $ w^i_1 w^i_2 ... w^i_{m_i})$ is generated by $\nu_i$). On the other hand, 
$|\varphi( u_1u_2 ... u_n/  w^i_1 w^i_2 ... w^i_{m_k})|  = \circ(n).$ 
 Hence, 
$|\varphi( u_1u_2 ... u_n/$ $  w^i_1$ $ w^i_2$ $  ...$ $ w^i_{m_i})| $  is minimal when $i = j$ (i.o., $u_1u_2 ... u_n$ is generated by $\nu_j$).
The theorem is proven.


\end{document}